\documentclass[11pt,a4]{article}

\usepackage[english]{babel}

\usepackage{epsfig}
\usepackage{dcolumn}
\usepackage{amsmath}
\usepackage{changebar}
\usepackage{graphicx}

\newcommand{\ba} {\begin{eqnarray}}
\newcommand{\ea} {\end{eqnarray}}
\usepackage{graphicx}
\usepackage{dcolumn} 
\usepackage{bm}      
\usepackage{umlaut}
\begin{document}
\renewcommand{\figurename}{Fig.}
\renewcommand{\tablename}{Tab.}
\title{Fluctuations of Particle Yield Ratios in Heavy-Ion Collisions}

\author{ A.~Tawfik\thanks{drtawfik@mti.edu.eg} \\
 {\small MTI Modern University, Faculty of Engineering, Cairo, Egypt }  
}

\date{}
\maketitle

\begin{abstract}
We study the dynamical fluctuations of various particle yield ratios at
 different incident energies. Assuming that the particle production yields in
 the hydronic final state are due to equilibrium chemical processes
 ($\gamma=1$), the experimental results available so far are compared with the
 hadron resonance gas model (HRG) taking into account the limited momentum
 acceptance in heavy-ion collisions experiments. Degenerated light and
 conserved strange quarks are presumed at all incident energies. At the SPS
 energies, the HRG with $\gamma=1$ 
 provides a good description for the measured dynamical 
 fluctuations in $(K^++K^-)/(\pi^++\pi^-)$. To reproduce the RHIC results,
 $\gamma$ should be larger than one. We also studied the dynamical
 fluctuations of $(p+\bar{p})/(\pi^++\pi^-)$. It is obvious that the
 energy-dependence of these dynamical fluctuations is non-monotonic.  
\end{abstract}

\section{\label{sec:A}Introduction}

Understanding of the dynamical properties of hot and dense matter is a key
question in the heavy-ion collisions experiments. The phase structure and
event-by-event fluctuations~\cite{shury,rajag,koch} have been 
suggested to provide comprehensive characteristics of the particle production
yields. They are essential observations to examine the hypothesis about the
equilibrium of the chemical processes in the hadron final
state~\cite{giorg}. The event-by-event fluctuations of certain particle yields
have been studied at SPS and RHIC
energies~\cite{Roland,NA491,STAR}. Therefore, it is natural to study the
energy-dependence of the particle yield ratios and 
the  event-by-event fluctuations using the hadron resonance gas model
(HRG), as it provided a good description for the thermodynamical evolution of
the hadronic system below the critical
temperature~\cite{Karsch:2003vd,Karsch:2003zq,Redlich:2004gp} and has been used
to characterize the conditions deriving the chemical
freeze-out~\cite{Taw3b,Taw3c}.

The hypothetical chiral symmetry breaking restoration and deconfined phase 
transition to quark-gluon plasma (QGP) are to be characterized by remarkable
fluctuations in the particle production yields~\cite{shury,rajag,koch}, which
are accompanied by dynamical and volume fluctuations as well. The latter can
simply be 
eliminated, when taking into consideration the dimensionless particle yield
ratios~\cite{koch}.   

In this letter, we try to answer the questions, whether the
strangeness quarks $q_s$ should enhance the dynamical fluctuations and whether
the critical endpoint could be localized by means of event-by-event
fluctuations in the hydronic final state. We make predictions for the
dynamical and statistical fluctuations of the ratios of different particle
yields in dependence on incident energy. Apparently, the dynamical
fluctuations strongly depend on sort of the particle yields. In some particle
yield ratios, the dynamical fluctuations are smaller than the statistical
ones. In others, the dynamical fluctuations are slightly greater than the
statistical ones. The energy dependence is non-monotonic.

All these predictions are phenomenologically of great interest and hope to
encourage experimental attempts to measure their event-by-event fluctuations
in a wide range of incident energy.

\section{\label{sec:2}Model}

The hadron resonances treated as a free
gas~\cite{Karsch:2003vd,Karsch:2003zq,Redlich:2004gp,Tawfik:2004sw,Taw3} are
conjectured to add to the thermodynamic pressure in the hadronic phase. This
statement is valid for free as well as strong interactions between the
hadron resonances 
themselves. Ii has been shown that the thermodynamics of strongly interacting
system can be approximated to an ideal gas composed of hadron
resonances~\cite{Tawfik:2004sw,Vunog}   

At finite temperature $T$, strangeness $\mu_S$ and baryo-chemical potential
$\mu_B$, the pressure of one sort of hadron resonance reads 
\begin{eqnarray}
\label{eq:lnz1} 
p(T,\mu_B,\mu_S) &=& \frac{g}{2\pi^2}T \int_{0}^{\infty}
           k^2 dk  \ln\left[1 \pm\,
           \gamma\,
           \lambda_B \lambda_S e^{\frac{-\varepsilon(k)}{T}}\right], 
\end{eqnarray}
where $\varepsilon(k)=(k^2+m^2)^{1/2}$ is single-particle energy and
$\pm$ stands for bosons and fermions, respectively. $g$ in the front of the
integration is the spin-isospin degeneracy factor and
$\gamma\equiv\gamma_q^n\gamma_s^m$ stand for the quark phase
space occupancy parameters, where $n$ and $m$ being number of light and
strange quarks, respectively. $\lambda=\exp(\mu/T)$ is the
fugacity, where $\mu$ is the chemical potential multiplied by corresponding
charge. Summing over all hadron resonances results in the final 
thermodynamic pressure in the hadronic phase, as no phase transition is 
conjectured in HRG. 

The quark chemistry is given by relating the {\it hadronic} chemical potentials
to each of the quark constituents; $\mu_B=3\mu_q$ and $\mu_S=\mu_q-\mu_s$,
where $q$ and $s$ being the light and strange quark quantum number,
respectively. The 
baryo-chemical potential for the light quarks is averaged as
$\mu_q=(\mu_u+\mu_d)/2$ and the strangeness chemical potential $\mu_S$ is
calculated as a function of $T$ and $\mu_B$ under the assumption of strange
quarks conservation~\cite{Tawfik:2004sw}.  \\

In grand canonical ensemble, the particle density is no longer
constant. 
\ba \label{eq:n1} 
\langle N\rangle &=& \sum_i \frac{g_i}{2\pi^2} \int dk k^2
\frac{\gamma e^{(\mu_i-\varepsilon_i)/T}}{1\pm \gamma e^{(\mu_i-\varepsilon_i)/T}}, \\
\label{eq:dn1} 
\langle (\Delta N)^2\rangle &=&  \sum_i \frac{g_i}{2\pi^2} \int dk k^2  
           \frac{\langle N_i\rangle}{1\pm \gamma e^{(\mu_i-\varepsilon_i)/T}}
\ea
To count the number density $N$ and fluctuations $(\Delta N)^2$ in the hadronic final state, the chemical
           freeze-out processes have to be 
taken into account, i.e. the hadron resonances should finally decay to stable
           particles or resonances. 
\ba \label{eq:n2}
\langle N_i^{final}\rangle &=& \langle N_i^{direct}\rangle + \sum_{j\neq
i} b_{j\rightarrow i} \langle N_j\rangle,\\ \label{eq:dn2} 
\langle (\Delta N_{j\rightarrow i})^2\rangle &=& b_{j\rightarrow i}
(1-b_{j\rightarrow 
i}) \langle N_j\rangle + b_{j\rightarrow i}^2 \langle (\Delta
N_{j})^2\rangle 
\ea
where $b_{j\rightarrow i}$ is the branching ratio for the decay of $j$-th
to $i$-th particle. In this work, the chemical freeze-out is characterized by
$s/T^3$~\cite{Taw3}, where $s$ is the entropy density. \\

The fluctuations of particle yield ratios of particle $1$ and particle $2$
read~\cite{koch} 
\ba  \label{eq:sigma}
\sigma^2_{N_1/N_2} &=& \frac{\langle (\Delta N_1)^2\rangle}{\langle
N_1\rangle^2} +  
                       \frac{\langle (\Delta N_2)^2\rangle}{\langle
                       N_2\rangle^2} -  
                     2 \frac{\langle \Delta N_1 \; \Delta
                     N_2\rangle}{\langle N_1\rangle \; \langle
                     N_2\rangle} 
\ea

In this expression, we include all possible fluctuations, i.e., dynamical and
statistical as well. The third term 
of Eq.~\ref{eq:sigma} counts for the fluctuations from hadron resonances that
decay into particle $1$ and particle $2$, simultaneously. In such a
mixing channel, all correlations including the quantum statistical ones are
taken into account. Obviously, this decay channel results in strong correlated
particles and correspondingly fluctuations. 

To extract the statistical fluctuations, we apply Poisson scaling to 
the mixed decay channels~\footnote[1]{Experimentally, there are various
  methods to eliminate the statistical fluctuations~\cite{STAR}. The
  frequently used one is the counting of particle yield ratios from mixing
  events.}, 
\ba \label{eq:sigmaStat}
(\sigma^2_{N_1/N_2})_{stat} &=& \frac{1}{\langle N_1\rangle} +
\frac{1}{\langle N_2\rangle} 
\ea
Detector acceptance factor and resolutions are main sources for statistical
fluctuations. Subtracting Eq.~\ref{eq:sigmaStat} from Eq.~\ref{eq:sigma}, we
get the dynamical fluctuations of the particle yield ratio $N_1/N_2$. 
\ba
 \label{eq:sigma2}
(\sigma^2_{N_1/N_2})_{dyn} &=& 
          \frac{\langle N_1^2\rangle}{\langle N_1\rangle^2} +
          \frac{\langle N_2^2\rangle}{\langle N_2\rangle^2} -
         \frac{\langle N_1\rangle+\langle N_2\rangle +
         2\langle N_1N_2\rangle}{\langle N_1\rangle\langle N_2\rangle}
\ea

\section{\label{sec:3}Results}

\begin{figure}[thb]
\centerline{
\includegraphics[width=12cm]{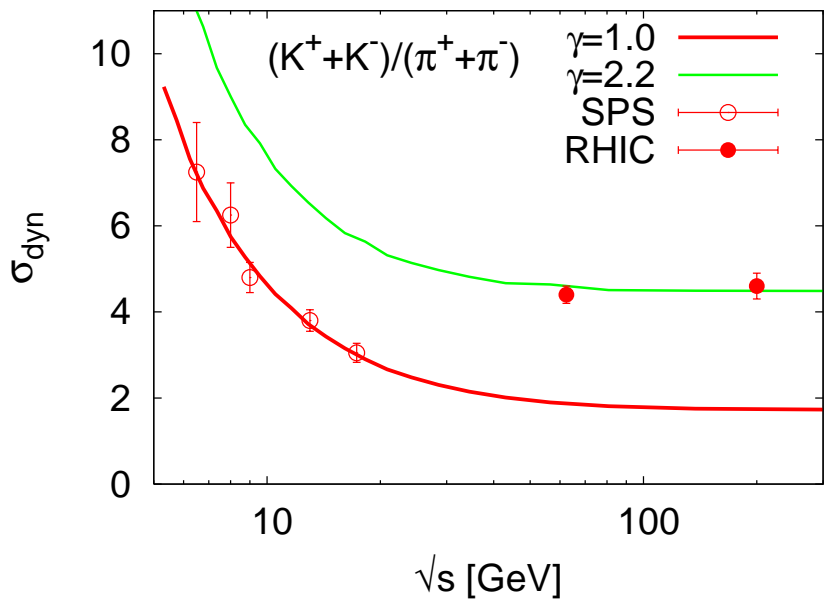}
}
\caption{The dynamical fluctuations of $(K^++K^-)/(\pi^++\pi^-)$ ratio as
 a function of square root of center of mass energy $\sqrt{s}$ (curves)
 compared with  experimental results
 (circles). SPS results~\cite{Roland}
 are drawn in open symbols while preliminary RHIC results~\cite{STAR} are
 given by solid circles. HRG with $\gamma=1.0$ can very well reproduce the SPS
 results. Larger values have to assign to $\gamma$ in order to
 reproduce the RHIC dynamical fluctuations (see the text). }   
 \label{Fig:kpi}   
\end{figure}

The experimentally measured dynamical fluctuations of particle yield 
ratios \hbox{$(K^++K^-)/(\pi^++\pi^-)$} are systematically confronted with the
theoretical predictions in 
Fig.~\ref{Fig:kpi}. An earlier attempt to compare with preliminary results 
has been reported in~\cite{koch,Roland}. It was found that the
theoretical and experimental ratios of dynamical to statistical fluctuations
are compatible with each other. Individual fluctuations themselves are not. 

\begin{figure}[thb]
\centerline{
\includegraphics[width=12cm]{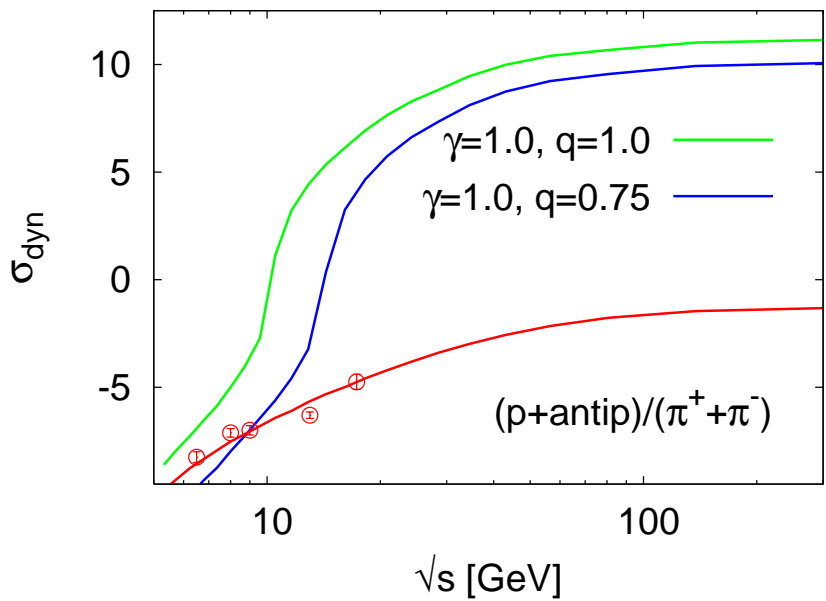}
}
\caption{The dynamical fluctuations of non-strangeness yield ratio
  $(p+\bar{p})/(\pi^++\pi^-)$. The experimental results (open circles) are
 taken from~\cite{Roland,NA491}. The negative values point to dominant
 statistical fluctuations. The factor $q$ relates canonical with grand
 canonical ensembles (see the text). }   
 \label{Fig:ppi}   
\end{figure}

Depending on $\gamma$, HRG is apparently able to predict various
particle yield ratios at a wide range of incident energy. 
At SPS energies, HRG with $\gamma=1$ provides a good description for the
experimentally measured dynamical fluctuations~\cite{Roland,STAR}. To reproduce
the RHIC results, $\gamma$ should be larger than one. The dynamical
fluctuations of \hbox{$(p+\bar{p})/(\pi^++\pi^-)$} are depicted in
Fig.~\ref{Fig:ppi}.  Few comments are in order at this moment. 
\begin{itemize}
\item The dependence on $\sqrt{s}$ is non-monotonic. The fluctuations can be
      suppressed and/or enhanced at different $\sqrt{s}$
\item Strangeness fluctuations are positive and enhanced with
      $\sqrt{s}$. There are remarkable minima at the top SPS
      energies 
\item At high energies, the fluctuations smoothly increase with $\sqrt{s}$ 
\end{itemize}

In Fig.~\ref{Fig:kpi}, we compare measured dynamical
fluctuations of $(K^++K^-)/(\pi^++\pi^-)$ yield ratios with HRG model. At
$\gamma=1.0$, we   
find an excellent agreement at SPS energies. At RHIC energy, the
measured fluctuations are above the theoretical ones. One has to
allow $\gamma$ to have values large than one, in order to reproduce the
data. The explanation for this disagreement would be two-fold. First, the RHIC
measurements are still preliminary~\cite{STAR}. The final
measurements might modify the $\gamma$-value reported here. Second, we refer to our previous study of the 
particle yield ratios in heavy-ion collisions~\cite{Taw1}, where we
concluded that the statistical models, like HRG, with $\gamma=1$ slightly
overestimate the particle yield ratios at RHIC energy. Therefore we should
assign to $\gamma$ values larger than one. What we observe here apparently
supports such a conclusion that the quark phase space occupancy factor,
characterized by $\gamma$, is obviously modified at RHIC energies. A
connection between $\gamma$ and hypothetical phase transition is discussed
in~\cite{Taw1}. Therefore, the modification of $\gamma$-value might be
understood, if phase transition to QGP takes place. 

Additionally, we see in Eq.~\ref{eq:sigma2} that $\langle N_i\rangle$,
where 
$i$ equals $1$ or $2$, inversely proportional to $\sigma$, assuming that the
contribution from the two-particle-channel is negligible $\langle
N_1N_2\rangle$. Should HRG model overestimate the particle yield ratios at RHIC
energy, as the case studied in~\cite{Taw1}, straightforwardly, we expect that
the dynamical fluctuations $\sigma$ should be underestimated at
$\gamma=1.0$. 

So far we can conclude that the quark phase space occupancy factor obviously
depends on the incident energy. Up to the top SPS energy, the quark phase
space occupancy factor is most probably saturated ($\gamma=1.0$). 
At RHIC energies, a phase transition to new state of matter has been
reported~\cite{Mclerran}. This can be seen in our analysis, if the quark phase
space occupancy factor is modified. Allowing $\gamma$ to have values larger
than one, HRG bests reproduce the particle yield ratios as reported
in~\cite{Taw3b,Taw3c,Taw1} and their dynamical fluctuations, the subject of
this work. \\   

The statistical model analysis of $(K^++K^-)/(\pi^++\pi^-)$ dynamical 
fluctuations reported in~\cite{Torr2} let to the conclusion that the
non-equilibrium model, equivalent to $\gamma\neq1$, would provide an
acceptable description at top SPS and RHIC energies. The definition of the
chemical freeze-out and the mechanism assuring conserved strangeness in the
hadronic final state would be responsible for this contradiction with HRG. 

At low incident energy, the energy density or temperature and the
degrees of freedom are not high enough to cause non-equilibrium phase
transition to QGP~\cite{Tawfik:2004sw,Taw3}. As the hadronic matter goes into
a new phase of nearly deconfined quarks and gluons, the quark phase space
occupancy factor should correspondingly change, The fitted values of $\gamma$
used to draw the figure 1 in~\cite{Torr2} are not explicitly given to discuss
them here. Taking into account the momentum acceptance, the HRG results show
that $\gamma=1$ best reproduces the dynamical fluctuations at low SPS
energies. \\

As the HRG calculations are performed in grand canonical formulism, the
so-called conservation laws~\cite{Gadz} would be taken into account. With
the conservation laws we mean - among others - the deference between the
canonical (CE) and grand canonical ensembles (GCE). At 
high energies, equivalent to large volume $V$, - in principle - there is 
almost no difference~\cite{Gadz} between CE and GCE. The conservation laws are
discussed at vanishing~\cite{Gadz} and finite chemical potential in the final
particle multiplicity~\cite{Gadz2}. 
In GCE, the control variables are $\mu$, $V$ and $T$. The total
number density $N$ is therefore allowed to fluctuate. Therefore $N$ is
related to CE by the Legendre transformation.  

In HRG, we studied this problem from another point of
view~\cite{Tawfik:2004sw} and found that the ideal quantum Boltzmann gas
formulism applied in HRG takes into account the interactions and 
the quantum statistics in the system. Furthermore, the ideal quantum Boltzmann
gas overcomes the limitations in classical Boltzmann ideal gas, that
the entropy for instance is only specified within an undetermined additive
constant in the way that it determines such an additive constant in the
high temperature limits of quantum Fermi- and Bose-gas.  
Quantitative estimation of the conservation laws
at finite chemical potential, i.e. different incident energies, is included in
the parameter $q$~\cite{Gadz}, where $q$ is probability to detect $n$
particle density out of total $N$ particle density produced in the whole
momentum space. The averaged particle density detected (accepted) read 
\ba
\left<n\right> &=& q  \left<N\right> \\
\left<(\Delta n)^2\right> &=& q^2  \left<(\Delta N)^2\right> + q  (1-q)
\left<N\right> 
\ea
where $\left<N\right>$ and  $\left<(\Delta N)^2\right>$ are given in
Eq.~\ref{eq:n1} - Eq.~\ref{eq:dn2}. \\

In Fig.~\ref{Fig:ppi}, the dynamical fluctuations of non-strangeness 
$(p+\bar{p})/(\pi^++\pi^-)$ yield ratios are depicted as
a function of $\sqrt{s}$. At $q=1$, HRG obviously overestimate
the dynamical fluctuations. It is worth noticing that $\sigma_{dyn}$ 
excursively increase at $\sqrt{s}\approx~15~$GeV.
The negative values 
are to be understood as dominant statistical fluctuations, especially the ones
not explicitly included in Eq.~\ref{eq:sigmaStat}. The negative values might
also refer to dominant fluctuations in the proton-pion channel.  

Taking into account an acceptance factor $q<1$, the experimental data can be
reproduced. As noticed in previous figure (Fig.~\ref{Fig:ppi}), the quark
phase space occupancy factor is
expected to be modified at incident energies higher than SPS. Taking this
fact into account, the dynamical fluctuations of $(p+\bar{p})/(\pi^++\pi^-)$
explosively switch to positive values. If this expectation turns to be correct,
we expect that the dynamical fluctuations of $(p+\bar{p})/(\pi^++\pi^-)$ at
RHIC energies might not be as smooth as the SPS ones.

\section{\label{sec:4}Discussions and conclusions}

At high temperatures, the hadronic matter is conjectured to go through 
chiral symmetry breaking restoration and deconfined phase transition(s) to
partonic phase at almost same critical
temperature~\cite{chiralT}. Depending on order of phase transition,
which in turn depends on quark flavors and masses, fluctuations in
particle yields are likely expected, especially if the transition causes 
out-of-thermal equilibrium. The three pions, $\pi^{0,\pm}$, are the lightest
Goldstone bosons resulting from chiral symmetry breaking. Therefore, strong
fluctuations in the pion fields are likely expected during the transition in
chiral symmetry. 
 
The chiral transition associated
with massless up and down quarks is of second order. If the quark masses are
accounted with, the transition is either a slightly first order or just a cross
over. The transition is explicitly a first order, if strange quark with very
light mass is included. Also the deconfinement phase transition depends on
the quark flavors and their masses. For three quark flavors, the transition
is likely of first order. In this case the transition occurs via nucleation
of hadronic bubbles in the background of QGP. For two flavors, the transition
is a smooth cross-over.  

This discussion might illustrate how strong are the pion dynamical
fluctuations connected with the chiral phase transition at high temperature
and out-of-thermal equilibrium. These fluctuations are given the generic name
of {\it disoriented chiral condensates} (DCC)~\cite{dcc}.

The dynamical fluctuations associated with strong first order of
phase transition are likely very large. The continues second order or cross
over phase transition might wish out large part of dynamical fluctuations in
the final state. On the other hand, the dynamical fluctuations are
conjectured to slow down near the second order phase transition. This has been
confirmed in classical systems, solid state physics. In quantum field theory,
the long-wavelength (spinodal) modes will be quenched through the second order
phase transition~\cite{quench}. \\ 

The Fluctuations of quark number $n_q$ have been studied in lattice
 QCD~\cite{karsch}. It has been found that the $T$-dependence
 of $n_q$ fluctuations is dominated by the analytic part of the partition
 function. Across the critical temperature, there is a smooth increase in the
 fluctuations. Almost same results are reported in this work. We therefore can
 use the fluctuations to characterize the phase transition.  

In our analysis, the critical temperature is not exactly specified, as HRG
 does implement any phase transition. The most economic way is to study the
 dynamical fluctuations in the hadronic final state, such a way we can also
 compare with the experimental results. We picked up two particle yield
 ratios, for them we 
 have extensive experimental estimations at a wide range of incident
 energies. We find that the dynamical fluctuations depend on the particle
yields and incident energy.

Tab.\ref{tab:1} gives the relation between dynamical and statistical
fluctuations $\sigma_{dyn}/\sigma_{stat}$ at SPS and RHIC energies. For the
strangeness particle yield ratios, the statistical fluctuations increase
against the dynamical ones.  For non-strangeness particle yield ratios, we get
just the opposite.

\begin{table}[htp]
\begin{center}
  \begin{tabular}{|c||c|c|c|c|c|}\hline
   $\sigma_{dyn}/\sigma_{stat}$    & $12.3$ & $17.3$ & $62.4$ & $200$ \\
   \hline\hline 
$(K+K^-)/(\pi^++\pi^-)$ & $0.0143$ & $0.0140$ & $0.0115$ & $0.0088$
   \\ \hline
$(p+\bar{p})/(\pi^++\pi^-)$ & $-0.0143$ & $0.0139$ & $0.0309$ & $0.0317$ \\
   \hline 
   \hline 
\end{tabular}
  \caption{\label{tab:1}$\sigma_{dyn}/\sigma_{stat}$ of 
 the particle yield ratios at $\gamma=1$ and $q=0.75$ and various
 $\sqrt{s}$. Using this quantity, 
 we can estimate how large are the dynamical fluctuations compared to
 the statistical ones when the produced particles are chemically frozen. The
 validity of these predictions apparently depend on the conditions
 controlling the chemical equilibrium in the final state. } 
\end{center}
  \end{table}

The energy dependence of dynamical fluctuations is non-monotonic. This can
also be seen in Tab.\ref{tab:1}. For
$(K+K^-)/(\pi^++\pi^-)$, the fluctuations decrease up
to top SPS energy, afterwards very slowly increase with
$\sqrt{s}$. $(p+\bar{p})/(\pi^++\pi^-)$ have negative dynamical fluctuations
at SPS energies. At higher energies, their dynamical fluctuations might jump to
positive values. The SPS energy might not high enough to lead to strongly
out-of equilibrium phase transition. The situation is different at RHIC
energies. On the other hand, the lattice QCD simulations show that the
transition in the region of the phase diagram corresponding to the RHIC energy
 - and very early universe - is a cross over. This continuous phase transition
 is not strong enough to secure out-of-equilibrium. Furthermore, any anomalous
 phenomenon, like the dynamical fluctuations, should be washed out in the final
 state, as the transition is a smooth and continuous cross-over or second
 order. This might be the reason why although new state of matter should be
 created at RHIC energies~\cite{Mclerran}, there is no abrupt change in the 
 dynamical fluctuations with increasing $\sqrt{s}$. That the dynamical
 fluctuations smoothly increase with $\sqrt{s}$ in agreement with the lattice
 simulations for $n_q$ fluctuations~\cite{karsch}, might support the
 conclusions that any anomalous phenomenon associated with non-equilibrium
 phase change likely would be washed out, if the phase change is smooth and
 does not cause out-of-equilibrium. \\

In our analysis, we assume that the particle production is due to chemical
equilibrium processes controlling the final state, i.e., $\gamma=1$. 
That our models can very well reproduce the experimental measurements means
that the equilibrium freeze out scenario is apparently correct,
especially up to SPS energies. Nevertheless, energy scan down to
$\sqrt{s}=10\;$GeV turns to be a crucial step to verify the worthwhile
behavior of particle production~\cite{Taw3,Taw1} and now dynamical
fluctuations.    

It is also worth extracting information about the role of
different decay channels in the energy-dependence of dynamical
fluctuations. It will be a further propose to study the effect of
chemical  non-equilibrium  processes on event-by-event dynamical
fluctuations.


\begin{thebibliography}{99}

\bibitem[1]{shury}E.~V.~Shuryak,~Phys.~Lett.~{\bf B}~423:9,~(1998)

\bibitem[2]{rajag} M.~A.~Stephanov, K.~Rajagopal, E.~V.~Shuryak,
           Phys.~Rev.~D~{\bf 60}~114028~(1999)

\bibitem[3]{koch}S.~Jeon, V.~Koch, Phys.~Rev.~Lett.~{\bf
           83}~5435~(1999)

\bibitem[4]{giorg}G.~Torrieri, S.~Jeon, J.~Rafelski,
           arXiv:nucl-th/0503026 

\bibitem[5]{Roland}Ch.~Roland [NA49 Collaboration],
  J.~Phys.~Conf.~Ser.~{\bf 27}~174~(2005);  
           J.~Phys.~G~{\bf 30}~S1381~(2004)

\bibitem[6]{NA491}S.~V.~Afanasiev, {\it et al.} [NA49 Collaboration],
           Phys.~Rev.~Lett~{\bf 86}~1965~(2000)

\bibitem[7]{STAR}S.~Das [STAR Collaboration], arXiv:nucl-ex/0503023

\bibitem[8]{Karsch:2003vd}F.~Karsch, K.~Redlich, A.~Tawfik,~
           Eur.~Phys.~J.~C~{\bf 29}~549,~(2003)

\bibitem[9]{Karsch:2003zq}F.~Karsch, K.~Redlich, A.~Tawfik,
           Phys.~Lett.~B~{\bf 571}~67,~(2003)

\bibitem[10]{Redlich:2004gp}K.~Redlich, F.~Karsch, A.~Tawfik,
           J.~Phys.,~G~{\bf 30}~S1271,~(2004) 

\bibitem[11]{Taw3b}A.~Tawfik, Nucl.~Phys.~A~{\bf 764}~387,~(2006)

\bibitem[12]{Taw3c}A.~Tawfik, Europhys.~Lett.~{\bf 75}~420~(2006);
           arXiv:hep-ph/0410392 

\bibitem[13]{Tawfik:2004sw} A.~Tawfik,~Phys.~Rev.~D~{\bf
           71}~054502,~(2005); arXiv:hep-ph/0412336

\bibitem[14]{Taw3}A.~TawfikJ.~Phys.~G~{\bf 31}:S1105,~(2005)

\bibitem[15]{Vunog}R.~Venugopalan, M.~Prakash, Nucl.~Phys.~A~{\bf 546}~718~(1992) 

\bibitem[16]{Taw1}A.~Tawfik and D.~Toublan, Phys.~Lett.~B~{\bf 623}~48-54~(2005)  

\bibitem[17]{Mclerran}M.~Gyulassy, L.~McLerran, Nucl.~Phys.~A~{\bf 750}~30~(2005)


\bibitem[18]{Torr2}G.~Torrieri, Int.~J.~Mod.~Phys.~E~{\bf 16}~1783-1789~(2007); Eur.Phys.~J.~C~{\bf 49}~287-292~(2007) 


\bibitem[19]{Gadz}V.~V.~Begun, {\it et al}, Phys.~Rev.~C~{\bf 70}~034901~(2004), arXiv:nucl-th/0404056

\bibitem[20]{Gadz2}V.~V.~Begun, {\it et al}, Phys.~Rev.~C~{\bf 76}~024902~(2007), arXiv:nucl-th/0611075 



\bibitem[21]{chiralT}F.~Karsch, E.~Laermann, Phys.~Rev.~D~{\bf 50}~6954~(1994)

\bibitem[22]{dcc} M.~Bleicher, {\it et.al.}, Phys.~Rev.~C~{\bf
    62}~041901~(2000); \\
S.~Gavin, J.~Kapusta, Phys.~Rev.~C~{\bf 65}~054910~(2002)

\bibitem[23]{quench} D.~Boyanovsky, H.~de~Vega,
   Phys.~Rev.~D~{65}~085083~(2002);\\ 
   M.~Pietroni, Phys.~Rev.~Lett.~{\bf 28}~2424~(1998)

\bibitem[24]{karsch} S.~Ejiri, F.~Karsch, K.~Redlich, Phys.~Lett.~B~{\bf
           633}~275,~(2006)








\end{thebibliography}
\end{document}